\documentclass[12pt]{article}
\usepackage{amsmath}    
\usepackage{color}      
\usepackage[]{graphicx}
\usepackage{indentfirst}
\usepackage{subfigure}
\title{Second harmonic generation in dissipative metamaterials}
\author{Ye.S. Mukhametkarimov$^{1}$, Zh.A. Kudyshev$^{1,2}$, A.E. Davletov$^{1}$, \\I.R. Gabitov$^{3,4}$, A.I.
Maimistov$^{5,6}$, M.G. Stepanov$^{3}$\\
\emph{$^1$Department of Physics, Al-Farabi Kazakh National University}\\
\emph{al-Farabi ave., 71, Almaty, 050038, KAZAKHSTAN}\\
\emph{$^2$Department of Electrical Engineering, University at Buffalo,}\\
\emph{State University of New York, 230 Davis Hall Buffalo, New}\\
\emph{York 14260-2500, USA}\\
\emph{$^3$Department of Mathematics, the University of Arizona,}\\
\emph{617 N. Santa Rita, Tucson, AZ 85721-0089, USA} \\
\emph{$^4$L.D. Landau Institute for Theoretical Physics, RAS,  1-A}\\
\emph{Akademik Semenova av., Chernogolovka, Moscow Region 142432,}\\
\emph{RUSSIA}\\
\emph{$^5$Department of Solid State Physics and Nanosystems,}\\
\emph{National research nuclear university MEPhI, Kashirskoe sh. 31,}\\
\emph{Moscow 115409, RUSSIA}\\
\emph{$^6$Department of Physics and Technology}\\
\emph{of Nanostructures, Moscow Institute for Physics and Technology,}\\
\emph{Institutskii lane 9, Dolgoprudny, Moscow region 141700,}\\
\emph{RUSSIA}}
\begin{document}
\maketitle
\begin{abstract}
Second harmonic generation is considered in lossy negative-index
metamaterials. It is shown that  energy transfer from fundamental to  harmonic takes place in the entire sample for the range of phase mismatch values. Note that in conventional case this range collapses to the point (ideal phase matching). The dependance of the boundary of this range as function of dissipation values is  obtained using computer simulations.
\end{abstract}

\maketitle

\section{Introduction}
In recent years  metamaterials have attracted a great deal of
attention in the scientific community. Research in this field is
stimulated by necessity for better understanding of fundamentals of
the electrodynamics in such materials and also because of
the  broad range of
potential applications\cite{veselago1968}-\cite{shelby2001}. One of
the most unusual type of metamaterials are those with a
negative refractive  index
(NRI)~\cite{veselago1968}, \cite{pendry2000}. The main difference of such
materials from conventional dielectrics is their left-hand orientation of
the fundamental triplet of vectors $\textbf{k, E}$ and $\textbf{H}$.
A consequence of such left-handed orientation is the opposite
directionality  of the wave vector $\textbf{k}$ and the Poynting
vector $\textbf{S}$ in NRI materials (NIRM).

Most of currently fabricated NRIM are utilizing plasmonic resonance
in metallic structures embedded to dielectric matrix. The sign of  the
index of refraction in this case is negative only for a limited frequency
domain. Nonlinear multi-wave interaction, when part of interacting
waves correspond to negative index frequency domain and another
part to positive index domain, is very different from conventional
multi-wave interaction. In particular case of second harmonic
generation, the propagation directions of fundamental and second
harmonics are opposite~\cite{popov2006}-\cite{scolara2006}.

In conventional nonlinear dielectrics, efficient energy transfer from
fundamental to  second harmonic wave takes place  only under
perfect phase matching. Intensities of fundamental and second
harmonics are respectively monotonically  decreasing and increasing through
the sample. Instead, in the presence of phase mismatch there is
alternating energy transfer between harmonics along the
sample~\cite{shen1984,armstrong1962} and field intensities  have
an oscillatory distribution along a sample.  It was been
shown~\cite{kudyshev2011}, that in NRI materials, monotonic energy
transfer takes place even in non-ideal phase matching conditions
$|\Delta| \ne 0$. Efficient energy transfer occurs within the
interval $|\Delta |\le \Delta_{cr}$, where $\Delta_{cr}$ is a
critical mismatch value. If the mismatch value is outside of the
critical interval, then field intensities are periodically varying
along the sample. These results were obtained under the assumption that
metamaterials are lossless. However, real metamaterails are
lossy~\cite{SCCYSDK05,XDKNCYS10} and loss values can be significant.
While the presence of losses in conventional materials does not change
dependance of spatial fields profiles on phase mismatch values,
losses may affect  the value of $\Delta_{cr}$ in negative index
materials.  This paper considers the process of the second harmonic
generation in the presence of losses. In particular, the impact of loss
values for both harmonics on the value of critical mismatch and
spatial distribution of field intensities are analyzed.

\section{Basic equations}

Following~\cite{kudyshev2011} we assume a refractive index that is
negative at  the fundamental frequency $\omega$ and  is positive
at the frequency of the second-harmonic wave $2\omega$. To satisfy
the phase matching condition, both wave-vectors  must be oriented
in the same direction. Therefore fundamental and the second
harmonic waves are propagating in opposite directions. The
propagation direction of the fundamental wave is assumed to be
oriented along the $z$ axes, while the propagation direction of
the second harmonic wave is oriented oppositely.

The set of equations in the slowly varying envelope approximation,
describing second harmonic generation in a lossy medium with
quadratic $\chi^{2}$
nonlinearity~\cite{armstrong1962,kudyshev2011} reads as:
\begin{equation}
    \begin{alignedat}{2}
        \frac{\partial E_{1}}{\partial z}&=-\imath
        \kappa_{1} E_{2}E_{1}^{*}\exp{(-\imath
        \Delta  z)}-\alpha_{1} E_{1},\\
        \frac{\partial E_{2}}{\partial z}&=\imath
        \kappa_{2} E_{1}^{2}\exp{(\imath \Delta z)}+\alpha_{2} E_{2}.
    \end{alignedat}
        \label{eq::main}
\end{equation}
Here $\Delta
=2k_1^{'}-k_2^{'}$ stands for the phase mismatch,
$k_{1,2}=k_{1,2}^{'}+\imath k_{1,2}^{''}$ are the wave numbers
of the fundamental and second harmonic correspondingly,   $E_{1,2}$ are
complex amplitudes of  fundamental and second harmonic waves
respectively,  $\kappa_{j} = 2
\pi\chi^{2}(\omega_{j})\omega_{j}^{2}\mu(\omega_{j})/\left(c^{2}
k^{'}_{j}\right),j=1,2$ denotes a coupling coefficients for
fundamental and second harmonic correspondingly, and
${\alpha}_{j}=k_{j}^{''}$ are the corresponding absorption
coefficients. The system of equations~(\ref{eq::main}) can be
transformed to autonomous form by the following change of variable
$E_{2}(z)\rightarrow E_{2}(z)\exp{(\imath \Delta  z)}$:
\begin{equation}
    \begin{alignedat}{2}
        \frac{\partial E_{1}}{\partial z}&=-\imath
        \kappa_{1} E_{2}E_{1}^{*}-\alpha_{1} E_{1},\\
        \frac{\partial E_{2}}{\partial z} &=\imath
        \kappa_{2} E_{1}^{2}-\imath \Delta  E_{2}+\alpha_{2} E_{2}.
    \end{alignedat}
        \label{eq::transformed}
\end{equation}

Let the left end of the sample coincide with the origin $z=0$ and
the right end correspond to the point $z=L$. Then the boundary
conditions for the set of equations \eqref{eq::main} have the
following form:
\begin{equation}
        E_{1}(0)=E_{10}\exp\left(i \varphi_{10}\right),~~E_{2}(L)=0.
   \label{eq::boundary}
\end{equation}
Here $e_{10}$, $\varphi_{10}$  are real amplitude and phase of the
incoming fundamental wave at the left end of the sample. Using
rescaling $E_1 =\sqrt{I_0}~{\cal E}_1(\zeta)$, $E_2
=\sqrt{I_0\kappa_{2}/\kappa_{1}}~{\cal E}_2(\zeta)$, $\zeta
=z\sqrt{\kappa_1 \kappa_2 I_0}$, $\widetilde{\Delta
}=\Delta/\sqrt{\kappa_1 \kappa_2 I_0}$, $\widetilde{\alpha
}_{1,2}=\alpha_{1,2}/\sqrt{\kappa_1 \kappa_2 I_0}$,  $I_0 =
E_{10}^2$, $l=L\sqrt{\kappa_1 \kappa_2 I_0}$ system of
equations~(\ref{eq::transformed}) can be rewritten as follows:
\begin{equation}
    \begin{alignedat}{3}
        \frac{\partial {\cal E}_{1}}{\partial \zeta}&=-\imath
       {\cal E}_{2}{\cal E}_{1}^{*}-\widetilde{\alpha}_{1} {\cal E}_{1},\\
        \frac{\partial {\cal E}_{2}}{\partial \zeta}& =\imath
         {\cal E}_{1}^{2}-\imath \widetilde{\Delta } {\cal E}_{2}+\widetilde{\alpha}_{2} {\cal E}_{2},\\
        {\cal E}_{1}(0) &=1,\quad {\cal E}_{2}(l)=0.
    \end{alignedat}
        \label{eq::transformed:dimensionless}
\end{equation}

It has been shown in~\cite{popov2006,kudyshev2011} that if
$\widetilde{\alpha}_{1,2}=0$, then the total energy flux is not
changing along the sample: $|{\cal E}_{1}|^2 -|{\cal E}_{2}|^2=C$.
This constant flux plays the role of the Manley-Rowe relation, which in
conventional dielectrics  represents conservation of energy
($|{\cal E}_{1}|^2 +|{\cal E}_{2}|^2=C$).  In lossy NRI materials,
equations for the field intensities read as
\begin{equation}
    \begin{alignedat}{2}
        \frac{\partial |{\cal E}_{1}|^2}{\partial \zeta}&=\imath
         \left({\cal E}_{1}^{2}{\cal E}_{2}^{*}-{\cal E}_{1}^{*2}{\cal E}_{2}\right)-2 \widetilde{\alpha}_{1} |{\cal E}_{1}|^{2},\\
        \frac{\partial |{\cal E}_{2}|^{2}}{\partial \zeta} &=\imath
         \left({\cal E}_{1}^{2}{\cal E}_{2}^{*}-{\cal E}_{1}^{*2}{\cal E}_{2}\right)+2\widetilde{\alpha}_{2} |{\cal E}_{2}|^{2}.
    \end{alignedat}
        \label{eq::intens}
\end{equation}
Therefore the total flux is not a constant and changes along the
sample in accordance to the following relation:
\begin{equation}
       \frac{\partial}{\partial \zeta}\left( |{\cal E}_{1}|^2 -|{\cal E}_{2}|^2
       \right)=-2 \left(\widetilde{\alpha}_{1}|{\cal E}_{1}|^2
       +\widetilde{\alpha}_{2}|{\cal E}_{2}|^2\right).
       \label{Manley:Rowe}
\end{equation}

This relation determines how the gradient of the total flux is related
to the energy dissipation per unit time in the unit of
volume and can be viewed as a differential form of Manley-Rowe
relation in the presence of losses.

By representing complex amplitudes ${\cal E}_{1}$ and ${\cal E}_{2}$
in terms of amplitudes $e_{1,2}$ and phases $\varphi_{1,2}$ and separation of the real and imaginary parts the following set of equations with corresponding boundary conditions can be obtained:
\begin{equation}
    \begin{alignedat}{4}
        &\frac{\partial e_{1}}{\partial \zeta} =e_{1}e_{2}\sin(\theta)-\widetilde{\alpha}_{1} e_{1},\\
        &\frac{\partial e_{2}}{\partial \zeta} = e_{1}^{2}\sin(\theta)+\widetilde{\alpha}_{2} e_{2},\\
        &\frac{\partial \theta}{\partial \zeta}=\left(\frac{e_{1}^{2}}{e_{2}}+2e_{2}\right)\cos{(\theta)}-\widetilde{\Delta}\\
        &e_{1}(0)=\exp\left(i
        \varphi_{10}\right),~~e_{2}(l)=0,~~\theta(l)=-\frac{\pi}{2}.
    \end{alignedat}
        \label{eq:real:main}
\end{equation}
here   $\theta =\varphi_2 -2 \varphi_1 $.

The boundary condition for $\theta$  can be found by taking into
account fact that ratio $e_{1}^{2}/e_{2}$ in the last equation of
the system~(\ref{eq:real:main}) is singular at the point $\zeta
=l$. Since the phase  derivative can not be infinite from the point of
view of physics, we conclude that $\cos\theta_{0} =0$ (here
$\theta (0)=\theta_{0}$) and $\theta_{0}=\pm \pi/2$. The slope of the
function $e_2 (\zeta)$ must be negative in the neighborhood of
$\zeta =l$, therefore, $\theta_0 =-\pi/2$.

Below, two important cases are considered:  ideal phase matching
condition $\widetilde{\Delta} = 0$ and second harmonic generation in
the presence of phase mismatch $\widetilde{\Delta}\ne 0$.

\section{Analysis and results of computer modeling}

\subsection{Ideal phase matching}

\begin{figure}[htbp]
    \centering

    \subfigure[Profiles intensity of the fields $e_{1,2}(\zeta)^2$ as function of  dimensionless coordinate $\zeta$. ]
    {
        \includegraphics[width=.6\columnwidth]{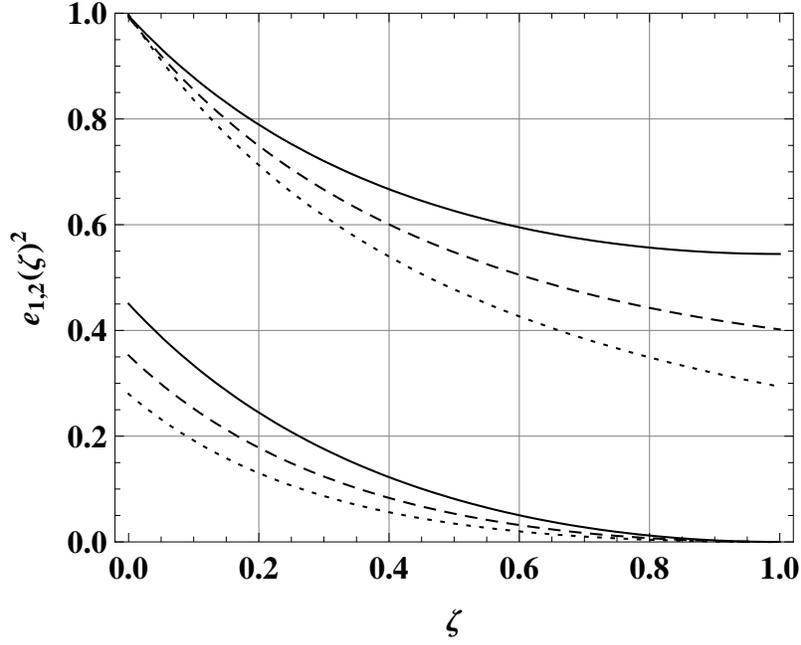}
        \label{fig::id::phase::a}
    }
    \subfigure[Total energy flux of the fields $e_{1}(\zeta)^2-e_{2}(\zeta)^2$ as function of  dimensionless coordinate $\zeta$.]
    {
        \includegraphics[width=.6\columnwidth]{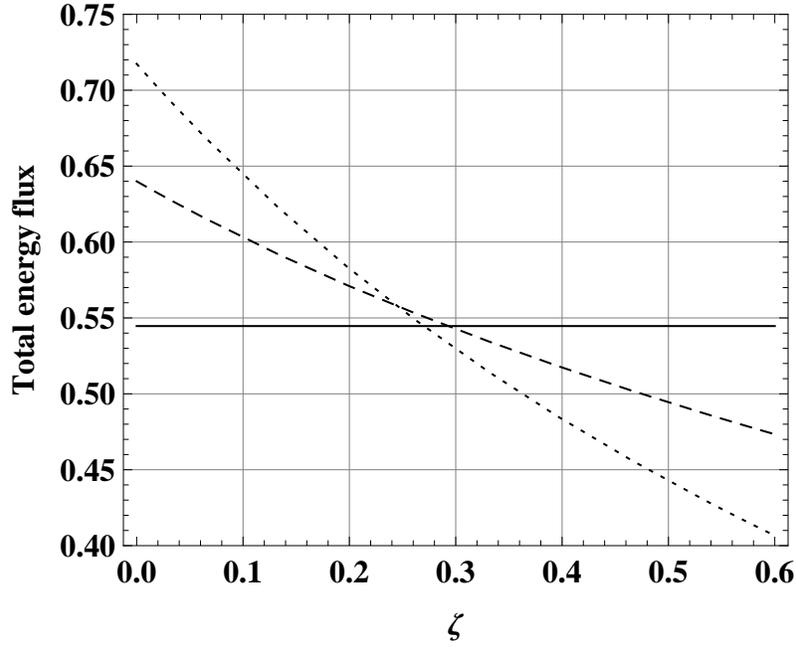}
        \label{fig:energy:flux}
    }
    \caption{The intensity dependence of the fundamental and the second-harmonic
    waves on the dimensionless coordinate along the sample under the condition
    of the ideal phase matching - subfigure (a) and spatial distribution  of the total energy flux
    $S=e_{1}^{2} - e_{2}^{2}$ - subfigure (b). Solid lines: $\tilde{\alpha}_{1,2} =0$,
    dashed lines: $\widetilde{\alpha}_{1} =0.2$, $\widetilde{\alpha}_{2} =0.1$, dotted-dashed lines:
    $\widetilde{\alpha}_{1} =0.4$, $\widetilde{\alpha}_{2} =0.1$.}
    \label{fig::id::phase}
\end{figure}

The phase difference $\theta (\zeta)$ is  not changing along the
sample and $\theta =-\pi/2$ in lossless metamaterials, when
$\widetilde{\Delta} =0$~\cite{popov2006,kudyshev2011}. Therefore
second harmonic generation can be described in terms of only two
equations for $e_{1,2}$. Presence of losses does not change the
equation for $\theta$, therefore $\theta =-\pi/2$ is still a stable
stationary solution of the equation for the phase $\theta$
in~(\ref{eq:real:main}). Indeed, since   $(e_{1}^2/e_{2}+2
e_{2})$ is always positive and $\widetilde{\Delta}=0$, then the
right-hand side of the equation for $\theta$ has negative slope at
the  points where $\theta =-(\pi/2)$. Hence, in the case of ideal phase
matching, the  phase difference is a constant  regardless of losses.
Losses affect only field intensities. Figure~\ref{fig::id::phase}
presents results from computer simulations and illustrates spatial
distribution of the filed intensities at ideal phase matching
$\widetilde{\Delta} =0$. Two principal cases are presented on
subfigure (a): the medium without and with loses. In the last case
values of the dimensionless absorption coefficients are chosen to
be $\widetilde{\alpha}_{1} =0.2$, $\widetilde{\alpha}_{2} =0.1$
and $\widetilde{\alpha}_{1} =0.4$, $\widetilde{\alpha}_{2} =0.1$.

Figure \ref{fig::id::phase}   clearly indicates that the presence of
the energy absorption in the medium does not alter the qualitative
picture of waves behavior inside the sample, and only contributes an
additional decrease in the intensities of the fundamental and the
second harmonic waves. It is quite natural that the decrease in the
intensities is aggravated by the growth of the corresponding
absorption coefficients.

The dependance of a total energy flux $S=e_{1}^{2} - e_{2}^{2}$ as
function of the coordinate is shown in subfigure (b). Solid line,
which corresponds to a lossless case, represents the Manley-Rowe
relation. Dotted-dashed and dashed lines in the insert are
solutions of the Manley-Rowe relation in differential
form~(\ref{Manley:Rowe}). Presence of losses leads to the increase
of the total flux value at the left end of the sample due to the
fact that losses are reducing inverse flux of second harmonic
energy. Total flux is monotonically decreasing along the sample
since spatial derivative of $S$ is  negative (see
equation~(\ref{Manley:Rowe})).
\begin{figure}[htbp]
    \centering
    \includegraphics[width=.5\columnwidth]{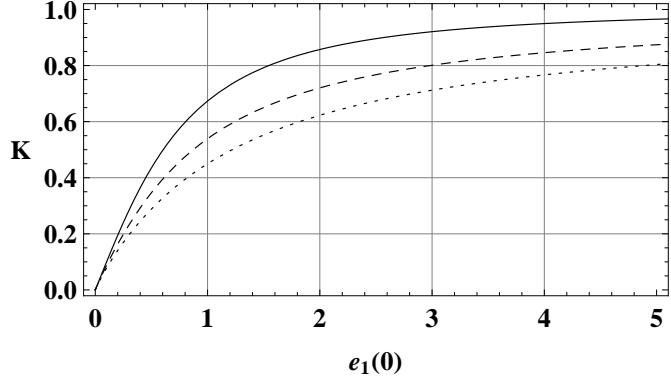}
    \caption{Conversion efficiency $K=e_{2}(0)/e_{1}(0)$ as a function
    of incident pump wave amplitude. Solid, dashed and dotted lines are
    corresponding to $\widetilde{\alpha}_{1,2}=0$, $\widetilde{\alpha}_{1,2}=0.3$, and
    $\widetilde{\alpha}_{1,2}=0.6$ respectively.} \label{conversion:efficiency}
\end{figure}
The impact  of losses on conversion efficiency $K=e_{2}(0)/e_{1}(0)$
is shown in Figure~(\ref{conversion:efficiency}). It was
demonstrated in~\cite{kudyshev2011}, that conversion efficiency in
lossless medium with ideal phase matching asymptotically
approaches $1$ (total conversion). Losses are slowing growth of
conversion efficiency with increase of the amplitude of  incident
pump wave and reducing its limit value.

\subsection{Impact of phase mismatch}

\subsubsection{Critical phase  mismatch in presence of losses.}

It has been shown in~\cite{kudyshev2011} that in NRI lossless
materials monotonic energy transfer  from pump to second harmonic
field takes place for  $|\widetilde{\Delta}|\le
\widetilde{\Delta}_{cr}$, where $\widetilde{\Delta}_{cr}$ is a
critical phase mismatch value. Outside of this interval both
fields exhibit spatial oscillatory behavior. Since losses are
unavoidable in realistic metamaterials,  it is  of practical
interest to analyze an impact of absorption in the medium on the
value of critical mismatch $\widetilde{\Delta}_{cr}$.

\begin{figure}[htbp]
    \centering

    \includegraphics[width=0.45 \textwidth]{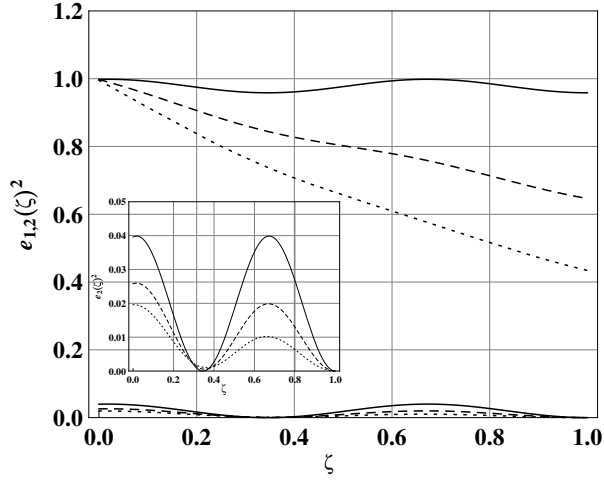}
    \caption{The intensities dependence of the fundamental
    and the second harmonic waves on the dimensionless coordinate
    at different values of the absorption coefficients $\widetilde{\alpha}_{1}$
    and $\widetilde{\alpha}_{2}$. The phase mismatch is fixed to $\widetilde{\Delta}=10$.
    Solid lines: $\widetilde{\alpha}_{1,2} =0$, dashed lines:
     $\widetilde{\alpha}_{1} =0.2$, $\widetilde{\alpha}_{2} =0.1$, dotted-dashed lines:
    $\widetilde{\alpha}_{1} =0.4$, $\widetilde{\alpha}_{2} =0.1$. The insert is an
    illustration of zoomed in behaviour of the second harmonic field $e_{2}(\zeta)$. }
    \label{fig::CriticDelta::10}
\end{figure}

Impact of losses on second harmonic generation is analyzed using
computer simulations.  Typical spatial profiles of both field
intensities  for $\widetilde{\Delta} = 10$ are shown in
Figure~\ref{fig::CriticDelta::10}. Solid lines correspond
to the case where $\widetilde{\alpha}_{1,2}=0$, dashed lines stand
for $\widetilde{\alpha}_{1} =0.2$, $\widetilde{\alpha}_{2} =0.1$
and the dotted-dashed lines displayed are for
$\widetilde{\alpha}_{1} =0.4$, $\widetilde{\alpha}_{2} =0.1$. It
is clearly seen from Figure \ref{fig::CriticDelta::10}
that, as in the case of the ideal phase matching, account of
losses in the medium does not qualitatively alter the behavior of
the wave fields within the sample as a function of the coordinate.
This fact strongly suggests that in the presence of dissipation
the two above mentioned regimes of the second harmonic generation
should remain intact.

\begin{figure}[htbp]
    \centering
    {
        \includegraphics[width=0.45 \textwidth]{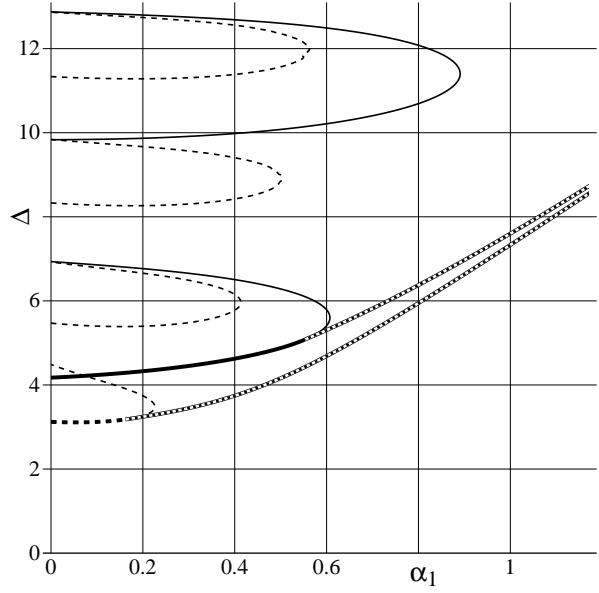}
    }
    \caption{The dependance of $\widetilde{\Delta}_{cr}$ as function of of  the absorption
    coefficients $\widetilde{\alpha}_{1}$ ($\widetilde{\alpha}_{2}=0$) for the given value of the
    incoming amplitude of the fundamental wave at the left end of the sample ${\cal E}_{1}(0) = 1$.
    Solid and dashed tongue-shaped curves lines are corresponding  to the solutions of equation $Q_{\widetilde{\Delta},l}(0) = 0$ (without constrain $Q_{\widetilde{\Delta}_{cr},l}(\zeta)\le 0$) for  $l=1$ and $l=2$  respectively. Part of the lowest branch of the lower curve in each case represents critical value  of the phase mismatch  $\widetilde{\Delta}_{cr} =\widetilde{\Delta}_{cr}(\widetilde{\alpha}_1)$. In case of $l=1$ this part (bold solid line) corresponds to the
     interval $0\leq \widetilde{\alpha}_1 \leq \overline{\alpha}_1$, here $\overline{\alpha}_1 \simeq 0.553$. All values of the function $Q_{\widetilde{\Delta},l}(\zeta)$ for $0\leq \widetilde{\alpha}_1 \leq \overline{\alpha}_1$ are satisfying to the constrain~(\ref{constrain}). Remaining part of
     this curve and all upper branches are irrelevant since $Q_{\widetilde{\Delta}, l}(\zeta) $
     changes sign inside $ 0<\zeta < l$.  Dependence of $\widetilde{\Delta}_{cr}(\widetilde{\alpha}_1)$ at
     $\widetilde{\alpha}_1 \geq \overline{\alpha}_1$ ($l=1$) is shown by dotted line. This line is tangential to
     the lower curve of the lowest branch at $\widetilde{\alpha}_1 =\overline{\alpha}_1$.  }
   \label{fig::out}
\end{figure}

Since the change in the character of the field distribution along
the sample from the monotonic to the oscillatory  regimes occurs
gradually, it is appropriate to adopt an explicit criterion for
identifying the critical value of the phase mismatch. It is
obvious that a direct application of that criterion for
$\widetilde{\alpha}_{1,2} = 0$ should reproduce the known
theoretical value of the critical mismatch
$|\widetilde{\Delta}_{cr}|=4 m_1$, here $m_1 \equiv e_{1}(l)$
(see~\cite{kudyshev2011}). This criterion can be introduced using
the system of equations~(\ref{eq::intens}). The first term in the
right hand side of~(\ref{eq::intens}) corresponds to the energy
exchange between fields. If this term is negative along the sample
then energy flows from pump to the second harmonic field. Positive
sign of this term corresponds to the inverse process when energy
flows from the second harmonic to the pump field. Therefore, the
function
\begin{equation}
       Q_{\widetilde{\Delta}, l}(\zeta)= {\cal E}_{1}^{2}{\cal E}_{2}^{*}-{\cal E}_{1}^{*2}{\cal E}_{2} = \frac{d |{\cal E}_{1}|^2}{d
\zeta} +2 \widetilde{\alpha}_{1} |{\cal E}_{1}|^2 = \frac{d |{\cal
E}_{2}|^2}{d \zeta} -2 \widetilde{\alpha}_{2} |{\cal E}_{2}|^2 ,
        \label{Exchange:criterion}
\end{equation}
can be used for searching $\widetilde{\Delta}_{cr}$. Note that
energy exchange between harmonics can also be characterized in
terms  of the angle $\theta$  in equations~(\ref{eq:real:main}).
The simplest case when
$\widetilde{\Delta}=\widetilde{\Delta}_{cr}$ corresponds to the
situation   when energy ``flows" from fundamental to second
harmonic in all points inside the sample except at $\zeta_{*}$ and
$\zeta =l$: $Q_{\widetilde{\Delta}_{cr},l}(\zeta)\le 0$ for $0\le\zeta
\le l$ and $Q_{\widetilde{\Delta}_{cr},l}(\zeta_{*})=0$,
$Q_{\widetilde{\Delta}_{cr},l}(l)=0$ (see
equation~(\ref{eq::intens})). The condition
$Q_{\widetilde{\Delta}_{cr},l}(l)=0$ is always valid since ${\cal
E}_{2}(l)=0$.  Therefore, the critical value of the phase mismatch
can be found by solving:
\begin{equation}
      Q_{\widetilde{\Delta}_{cr},l}(\zeta_{*})=0,        \label{Equation:Q}
\end{equation}
with the constrain:
\begin{equation}
Q_{\widetilde{\Delta}_{cr},l}(\zeta)\le 0. \label{constrain}
\end{equation}
Figure~\ref{fig::out} represents phase mismatch
$\widetilde{\Delta}$  satisfying equation~(\ref{Equation:Q}) as a
function of $\widetilde{\alpha}_1$.
Equation~(\ref{Equation:Q}) was solved using Newton's method. Both
fields ${\cal E}_{1,2}(\zeta)$ were found by solving the system  of
equations~(\ref{eq::transformed:dimensionless}) assuming that
$\widetilde{\alpha}_2 =0$. The sign of the refraction index
corresponding to the  frequency of  fundamental  harmonic is
negative. In most cases a negative sign of refractive index is
achieved using plasmonic resonance in the metallic structures,
which leads to considerable losses. Therefore it reasonable to
assume   that losses at the second-harmonic frequency are much
smaller  than the losses on the frequency of the fundamental
harmonic.

We considered two cases: when length of the  sample is chosen to be
$l = 1$ (solid line) and $l = 2$ (dashed line). The amplitude of
the incident pump wave in both cases is chosen to be ${\cal
E}_{1}(0) = 1$. Numerical simulations showed that the value of critical
mismatch is increasing with $\alpha_1$.  It also follows from computer
simulations that $\zeta_{*}=0$  for values of $\widetilde{\alpha}_1$
in the interval $0\le \widetilde{\alpha}_1 \le \overline{\alpha}_{1}$.
In other words, equation~(\ref{Equation:Q}) in this interval takes
the form: $Q_{\widetilde{\Delta}_{cr},l}(0)=0$. Branches on
Figure~\ref{fig::out} correspond to
the  multi-valued solutions of the implicit equation
$Q_{\widetilde{\Delta},l}(0)=0$ for different values of $\alpha_1$.

\begin{figure}[htbp]
    \centering
    \includegraphics[width=0.45 \textwidth]{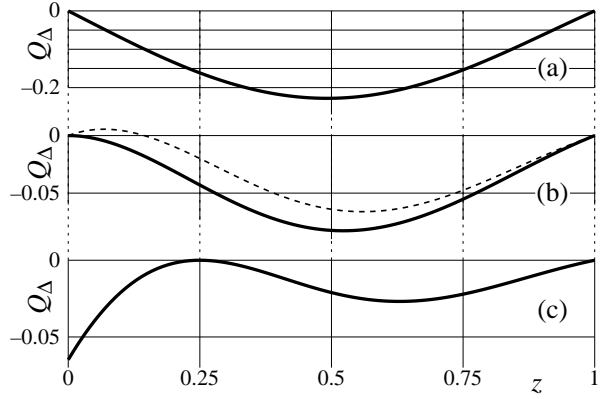}
    \caption{Spatial profiles of the function  $Q_{\widetilde{\Delta}}(\zeta)$ corresponding to different
    branches of the solid line on Fig.~(\ref{fig::out}), $l = 1$.
    a) $\widetilde{\alpha}_1 = 0.2$;
    b) solid line  $\widetilde{\alpha}_1\approx 0.553$ and dashed line $\widetilde{\alpha}_1 = 0.6$;
    c) $\widetilde{\alpha}_1 \approx 0.814$. In this case, there is maximum at point $\zeta^* = 0.25$, where $Q_{\widetilde{\Delta}} = 0$.}
   \label{fig::qz}
\end{figure}

Part of the lowest branches (solid bold line $l=1$  or solid
dashed line $l=2$)  corresponds to the dependance of the critical
value $\widetilde{\Delta}_{cr}$ on $\widetilde{\alpha}_1$. In this
case $Q_{\widetilde{\Delta},l}(\zeta)<0$ for $0<\zeta< l$. An example
of such function   $Q_{\widetilde{\Delta}, l}(\zeta) $ for
parameters $\widetilde{\alpha}_1 = 0.2$, $l=1$ is shown in
Figure~(\ref{fig::qz}), insert (a).

Without loss of generality, we consider the case when $l = 1$. Our
analysis shows that the function $Q_{\widetilde{\Delta},l}(\zeta)$ is
zero at $\zeta =0$ together with its first  derivative
$\left(Q_{\widetilde{\Delta},l}(0)\right)^{\prime}_{\zeta}=0$
when $\widetilde{\alpha}_1 =\overline{\alpha}_1$;
here $\overline{\alpha}_1 \simeq 0.553$. The profile of the function
$Q_{\widetilde{\Delta}, l}(\zeta) $ in this case is shown in
Figure~(\ref{fig::qz}),  solid line in insert (b).

The remaining part of the lowest  branch  corresponds to the case
when $Q_{\widetilde{\Delta},l} (\zeta)$ has one zero inside the
interval $0<\zeta<l$. The profile of such function is shown in
Fig.~(\ref{fig::qz}), dashed line in the insert (b). This function
corresponds to a $\widetilde{\Delta}$ which is obtained from equation
the equation $Q_{\widetilde{\Delta},l}(0)=0$ for
$\widetilde{\alpha}_1 > \overline{\alpha}_1$.
Note that in this case, the  constrain
$Q_{\widetilde{\Delta}_{cr},l}(\zeta)\le 0$ is not valid, therefore
the corresponding $\widetilde{\Delta}$ does not belong to the family of
critical values.  The values of $\widetilde{\Delta}$ on the upper branch
correspond to functions of $Q_{\widetilde{\Delta},l} (\zeta)$ with several
zeros and also do not belong to the family of critical values of
phase mismatch.

The dotted  line in Figure~(\ref{fig::out}) shows dependance of
critical mismatch on $\widetilde{\alpha}_1$ for
$\widetilde{\alpha}_1 \geq \overline{\alpha}_1$ ($l=1$). The function
$Q_{\widetilde{\Delta},l} (\zeta)$ corresponding to this case is
negative and has one zero within the interval $0\leq \zeta \leq l$
at $\zeta =\zeta_{*}$. The derivative of  this function with respect
to $\zeta$ at zero point is also equal to  zero:
\begin{equation}
       Q_{\widetilde{\Delta}, l}(\zeta_{*}) = 0,\quad \frac{d Q_{\widetilde{\Delta}, l}(\zeta_{*})}{ d \zeta}= 0
        \label{Q_Qz}
\end{equation}
An example of such function is shown in Figure~(\ref{fig::qz}),
insert (c).  The bold dotted curve is tangential to the lowest
branch at the point $\widetilde{\alpha}_1 \simeq \overline{\alpha}_1$.
Finally, the critical value of mismatch as a function of
$\widetilde{\alpha}_1$ is shown as a bold curve for
$\widetilde{\alpha}_1 \le \overline{\alpha}_1$ and
as a bold dotted curve for
$\widetilde{\alpha}_1 > \overline{\alpha}_1$ ($l=1$).

\begin{figure}[htbp]
    \centering
    \includegraphics[width=0.45 \textwidth]{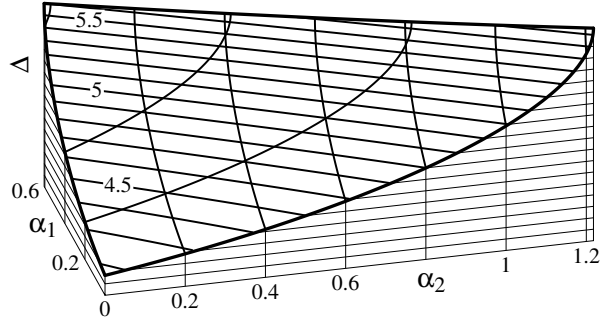}
    \caption{The dependence of the critical phase mismatch $\widetilde{\Delta}_{cr}$ on the absorption
    coefficients $\widetilde{\alpha}_{1}$ and $\widetilde{\alpha}_{2}$ for the given value of the
    incoming amplitude of the fundamental wave at the left end of the sample ${\cal E}_{1}(0)
= 1$.}
   \label{fig::3D2}
\end{figure}

The dependance of the critical mismatch $\widetilde{\Delta}_{cr}$
on both absorption coefficients $\widetilde{\alpha}_{1,2}$ can be
found in a similar way. In  case of two variables
$\widetilde{\alpha}_{1}$ and $\widetilde{\alpha}_{2}$ each branch
(see Fig.~\ref{fig::out} ) will span the corresponding surface. The
behavior of $\widetilde{\Delta}$, satisfying the
equation~(\ref{Equation:Q}), as function of $\widetilde{\alpha
}_{1,2}$ is shown on Fig.~\ref{fig::3D2}. This figure portrays a two
dimensional generalization of the lower part of the lowest branch
for $l=1$ and ${\cal E}_{1}(0) = 1$ shown on Fig.~\ref{fig::out}.
Similarly to the case which is considered above, only part of this
surface represents  critical values of phase mismatch. The domain of
critical values ${\widetilde\Delta}_{cr}$ can be found by means of
imposing the additional condition
$\left(Q_{\widetilde{\Delta},l}(0)\right)^{\prime}_{\zeta}=0$.
Outside of this domain $\widetilde{\Delta}_{cr}$ can be found by
solving the system of equations~(\ref{Q_Qz}).

\subsubsection{Field profiles in presence of losses}

\begin{figure}[htbp]

    \subfigure[$~$ Intensity and phase profiles in the  ideal case: $\widetilde{\alpha}_{1,2}=0$.]
    {
        \includegraphics[width=0.45 \textwidth]{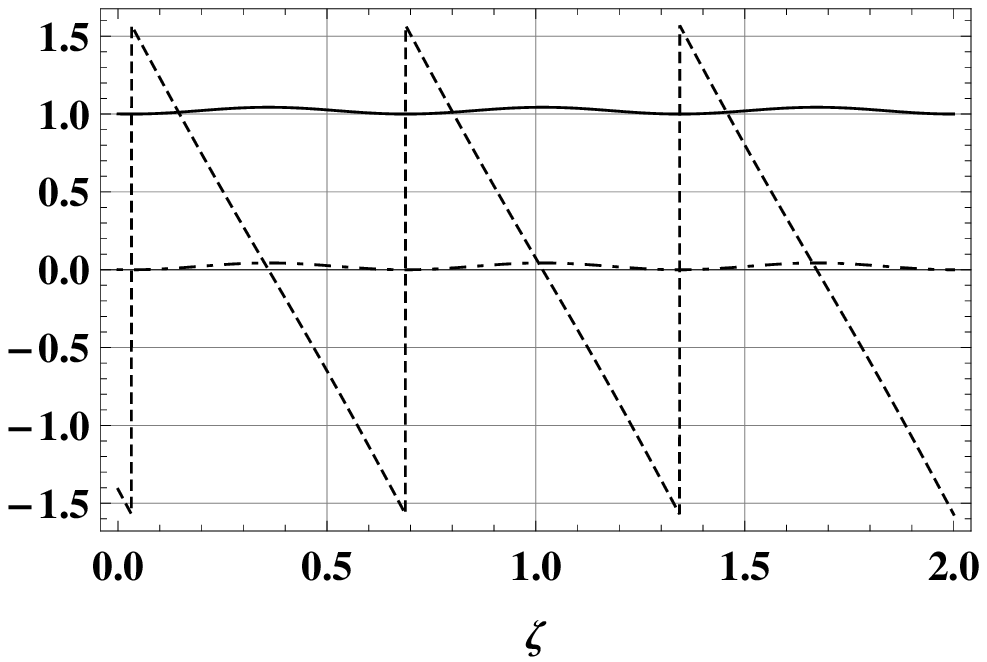}
        \label{profiles:ideal:case}
    }
    \subfigure[$~$ Intensity and phase profiles in presence of losses: $\widetilde{\alpha}_{1,2}=0.1$.]
    {
        \includegraphics[width=0.45 \textwidth]{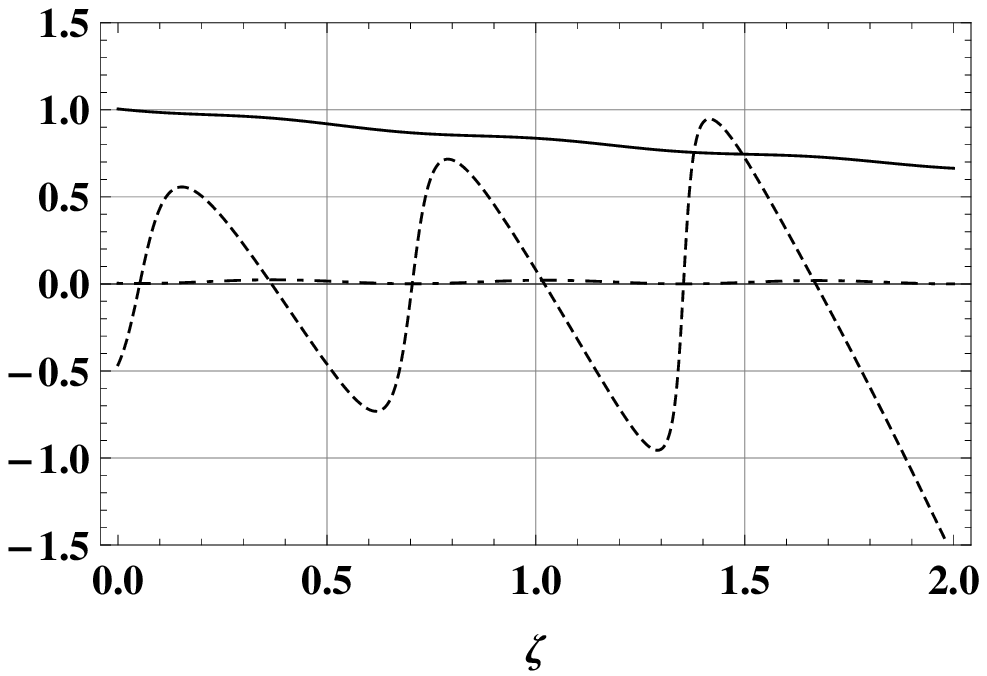}
        \label{profiles:ideal:losses}
    }
    \caption{Spatial profiles corresponding to supercritical regime  $\widetilde{\Delta} =10$ for
    field intensities $e_{1,2}^2$ and phase $\theta$ in the ideal case (left subfigure)
    $\widetilde{\alpha}_{1,2}=0$ and in presence of losses (right subfigure) $\widetilde{\alpha}_{1,2}=0.1$. Pump
    intensity, second harmonic intensity   and phase $\theta $ are
     labeled as $1$, $2$ and  $3$ respectively.}
    \label{Profiles:general}
\end{figure}

Second harmonic generation in the subcritical case is similar to the
case of ideal phase matching described above. This subsection presents
results of computer simulations describing the phase and field
profiles along the sample in the supercritical regime.
Figure~\ref{Profiles:general} shows an example of spatial profiles
corresponding to this regime $\widetilde{\Delta} =10$ for
both field intensities $e_{1,2}^2$ and phase $\theta$ in the ideal
case (Fig.\ref{profiles:ideal:case}) and in the presence of losses
(Fig.~\ref{profiles:ideal:losses}). Fig.~\ref{profiles:ideal:case}
portrays periodic intensity oscillations corresponding to
alternating energy exchange between pump and second harmonic
fields. Each time the amplitude of second harmonic ``touches"
zero $e_2 =0$, the phase $\theta$ experiences ``$\pi$-phase slip",
similar to a phase slip observed in~\cite{GKC94}.  Presence of
losses leads to a smoothing of this phase jump, which is shown in
Fig.~\ref{profiles:ideal:losses}. Note that the value of the phase
$\theta$ at the end of the sample is $\theta (l)=-\pi/2$ in both
cases. Since $\cos(-\pi/2)=0$, such value of $\theta$ eliminates
singularity at the right hand side of
equation~(\ref{eq:real:main}) at the end of the sample $\zeta=l$
where $e_2 (l)=0$ and is consistent with negative sign of the
derivative $e_2^{\prime} (\zeta)< 0$ near $\zeta =l$. The intensity
profile of the  pump field shown in Fig.~\ref{profiles:ideal:case}
indicates that at the points of maxima, the  values of the  pump
field intensities are greater than intensity of the incident pump
field $e_1^2 >e_0^2$. This observation does not contradict
conservation of energy. It should be noted that these solutions
are representing stationary equilibrium states describing the
interaction of two opposite waves. Interaction of these  waves in
the supercritical regime leads to spatial segmentation of the
interval $[0,l]$ to alternating subintervals in which  energy
flows (in the spectral domain) from fundamental harmonics to second
harmonic field and in the next subinterval energy flow changes its
direction. Energy flow from fundamental to second harmonics takes
place when phase is positive $\theta >0$. Energy flow in opposite
direction takes place when phase is negative $\theta <0$. As an
example let us  consider the point $\zeta =l$, where $e_2 (l)=0$.
The field $e_2 (\zeta)$ is growing while it propagates from right
to left, therefore $e_1 (\zeta)$ is loosing energy and
decreasing while it propagates in $\zeta$ ``direction". The fact
that  $e_{1}(0)$ is less than the maximal value of
$e_{1}$ inside the sample means that in the neighborhood of the
point $\zeta =0$ energy "flows" from fundamental to second
harmonic.

\begin{figure}[htbp]
    \centering
    \includegraphics[width=0.45 \textwidth]{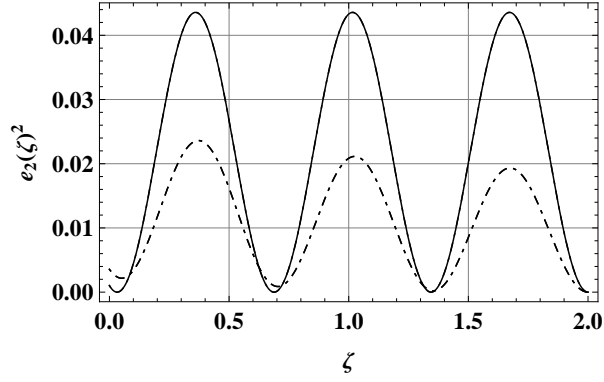}
    \caption{Profiles of intensity second harmonic in presence of losses
    $\widetilde{\alpha}_{1.2}=0.1$ (dashed line) and lossless case
    $\widetilde{\alpha}_{1.2}=0$ (solid line)  with $\widetilde{\Delta}=10$, $e_1(0)=1$.}
    \label{second:harm}
\end{figure}

The difference in the behavior of second harmonic intensities
without and with losses is illustrated in Fig.~\ref{second:harm}.
In the ideal case (solid line) minimal value of the second
harmonic intensity is zero $\min e_2^2=0$ and $e_2$
periodically oscillates along the sample. In the presence of losses
$e_2 (\zeta)$ is not periodic anymore,  $e_2=0$ holds only at the
end point $\zeta=l$ and minimal values of $e_2$ are small
$e_2(\zeta_{\min})\ll e_1(\zeta_{\min})$ and increasing from the
right to the left. Losses are regularizing sharp $\pi-$ phase
slips, which becomes smooth, wider and smaller than $\pi$.  Note
that a rapid change of $\theta (\zeta)$ takes place in the vicinity
of the local minima of $e_2(\zeta_{min})$.
Fig.~\ref{profiles:ideal:losses} clearly indicates the presence of two
scales: a fast scale of the change of $\theta(\zeta)$ near $\zeta \sim
\zeta_{min}$ (positive slope) and a slow change of $\theta(\zeta) $
(negative slope). The slow scale dynamics is determined by the
second term  $\widetilde{\Delta}$ in the right-hand side of the
equation for $\theta$ of the system~(\ref{eq:real:main}). The fast
dynamics is determined by the first term  of the right hand side
when $e_2$ becomes small $\varepsilon=e_2(\zeta_{\min}) \ll 1$.
In the leading order behavior of the ``regularized" phase slip the
second harmonic field reads as

\begin{eqnarray}
 e_2(\zeta) \simeq \sqrt{\varepsilon^2 + e_1^4 (\zeta - \zeta_{min})^2} \label{e2:theor} \\
\sin \theta(\zeta) \simeq \frac{e_1^2 (\zeta - \zeta_{min})}{\sqrt{\varepsilon^2 + e_1^4 (\zeta - \zeta_{min})^2}}\label{sin:theor}
\end{eqnarray}

\begin{figure}[htbp]
    \centering

    \subfigure[Comparison of the profiles for $e_2(\zeta)$ obtained by computer modeling
    -- dashed line, and by the equation~(\ref{e2:theor}) -- solid line. As seen from the figure the two curves coincide to the line width. ]
    {
        \includegraphics[width=0.45 \textwidth]{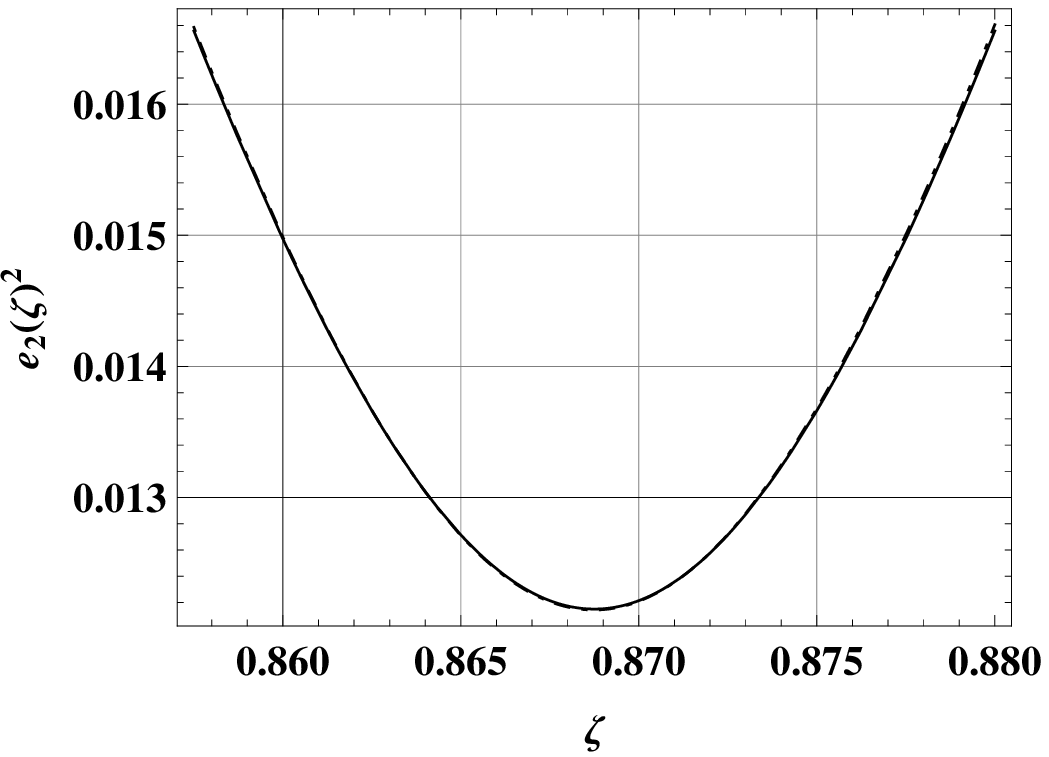}
        \label{profiles:e2}
    }
    \subfigure[Profiles of the fields $e_{1,2}(\zeta)$ and $\sin\theta(\zeta)$ obtained by
    computer modeling -- large-dashed, small-dashed, and dot-dashed lines respectively.
    Solid line represents $\sin\theta(\zeta)$ given by the equation~(\ref{sin:theor}).  ]
    {
        \includegraphics[width=0.45 \textwidth]{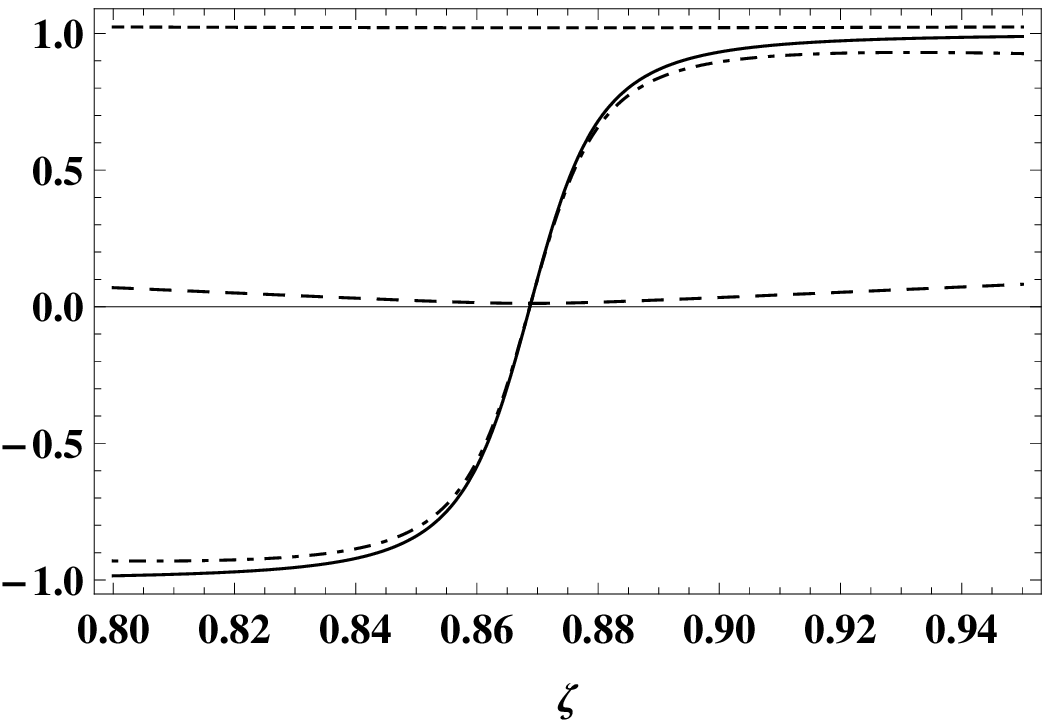}
        \label{profiles:phase}
    }
    \caption{Profiles of $e_{1,2}(\zeta)$ and $\sin\theta$ obtained by computer  and given by
    equations~(\ref{e2:theor}) - (\ref{sin:theor}), here $\widetilde{\Delta} =6$, $\widetilde{\alpha}_{1,2}=0.005$,
    $\varepsilon \approx 0.01215$, $\zeta_{min} \approx 0.86875$.}
    \label{Profiles:comparizon}
\end{figure}

Comparison of the results obtained by direct computer modeling and
from equations~(\ref{e2:theor})-(\ref{sin:theor})  is shown in
Fig.~\ref{Profiles:comparizon}. Minimum point $\zeta_{min}
\approx 0.86875$ and  values of $\varepsilon
=e_2(\zeta_{min})\approx 0.01215$, and $e_1(\zeta_{min})\approx1$
are taken from the results of computer simulations for $\Delta =6$,
$\alpha_{1,2}=0.005$ and substituted in
equations~(\ref{e2:theor})-(\ref{sin:theor}).
Fig.~\ref{Profiles:comparizon} shows that
equations~(\ref{e2:theor})-(\ref{sin:theor}) are describing well the
``fast" scale supercritical dynamics of the
system~(\ref{eq:real:main}).

\section{Conclusion}

The process of second harmonic generation in dissipative
metamaterials has been studied  in  case of  ideal and non-ideal
phase matching. Similarly to  lossless medium, the
existence of two regimes of  second harmonic generation  was demonstrated
theoretically. One regime corresponds to ``unidirectional" energy transfer
from fundamental to second harmonic and results in monotonic behavior of
the field profiles along
the  sample. Another regime occurs at higher values of the phase
mismatch and corresponds to the case when energy flow changes ``direction"
and leads to oscillatory field profiles along the sample. The critical
phase mismatch $\Delta_{cr}$, separating these regimes,
depends on the length of the sample and  on absorption
coefficients of both waves.    Analysis of the
second harmonic generation  in the oscillatory regime shows the difference in
behaviour  of the electric fields phase difference
for  the ideal and lossy cases. When phase mismatch value is larger than
$\Delta_{cr}$, the phase difference experiences $\pi$-phase slip.
Presence of losses give a smoothing mechanism of the phase jump and
reduces jump's amplitude

\section{Acknowledgments}

We would like to thank V. P. Drachev for valuable discussions.
AIM, YeM and  ZhK appreciate the support and hospitality of the
University of Arizona Department of Mathematics during the
preparation on this manuscript. This work was partially supported
by NSF (Grant No. DMS-0509589), ARO-MURI Award No. 50342-PH-MUR
and the State of Arizona (Proposition 301), RFBR (Grants No.
09-02-00701-a and No. 12-02-00561),  the Federal Goal-Oriented
Program "Scientific and Scientific-Educational Personnel of
Innovational Russia agreement 8834" and  the Ministry Of Sciences
and Education of Kazakhstan  grant GF3 No$/$1567.

\end{document}